\documentclass[runningheads]{llncs}
\usepackage{graphicx}
\usepackage{enumerate}
\usepackage{todonotes}
\usepackage{floatrow}
\usepackage{float}

\newfloatcommand{capbtabbox}{table}[][\FBwidth]
\begin{document}
%
\title{Man-in-the-OBD: A modular, protocol agnostic firewall for automotive dongles to enhance privacy and security \thanks{The authors acknowledge the financial support by the Federal Ministry of Education and Research of Germany in the program of "Souverän. Digital. Vernetzt." Joint project 6G-RIC, project identification number: 16KISK034.}}
\titlerunning{Man-in-the-OBD}
%
 \author{Felix Klement\inst{1}\orcidID{0000-0001-9650-7698}  \and
 Henrich C. Pöhls\inst{2}\orcidID{0000-0002-7256-0387} \and
 Stefan Katzenbeisser\inst{1}}
 \authorrunning{Klement et al.}
 \institute{Chair of Computer Engineering, University of Passau, Germany
 \email{\{felix.klement,stefan.katzenbeisser\}@uni-passau.de}
 \and Chair of IT Security, University of Passau, Germany
 \email{\{hp\}@sec.uni-passau.de}}

\maketitle              
\begin{abstract}
Third-party dongles for cars, e.g. from insurance companies, can extract sensitive data and even send commands to the car via the standardized OBD-II interface. Due to the lack of message authentication mechanisms, this leads to major security vulnerabilities for example regarding the connection with malicious devices. Therefore, we apply a modular, protocol-independent firewall approach by placing a man-in-the-middle between the third-party dongle and the car's OBD-II interface. With this privileged network position, we demonstrate how the data flow accessible through the OBD-II interface can be modified or restricted. We can modify the messages’ contents or delay the arrival of messages by using our fine-granular configurable rewriting rules, specifically designed to work protocol agnostic.  We have implemented our modular approach for a configurable firewall at the OBD-II interface and successfully tested it against third-party dongles available on the market. Thus, our approach enables a security layer to enhance automotive privacy and security  of dongle users, which is of high relevance due to missing message authentications on the level of the electronic control units.

\end{abstract}
\section{Introduction}
\label{motivation}

Today's cars, self-driving or not,  can exchange information with the outside over a standardised interface: the On-Board-Diagnosis (OBD) interface, which exists since 1988. OBD-II is a vehicle diagnostic system, which makes important electronic control units (ECU) of the car and their data accessible. This allows reading the current speed, rpm and other information from the car. In 1996 the USA made it mandatory for every new car that is sold to have this interface. 
According to Regulation (EC) No. 715/2007, all new passenger car registrations in the EU since 2001 for gasoline engines and since 2004 also for diesel engines must be equipped with an OBD interface. 
This means that nowadays a huge amount of cars are equipped with this On-Board-Diagnosis system.  
However, this also means that the driver is strongly encouraged or forced to connect external devices (dongles) to this port first, so that they can then read out the car's data via the OBD interface.
In this paper we show that the OBD interface lacks basic data authentication mechanisms making it possible to place our firewall as a man-in-the-middle. 
In the remainder of this paper we will use the term \textit{dongle} to refer to the third-party device that will be connected to the car over the OBD-II interface. 
%
Dongles might offer all functionality on their own (standalone) or consisting of a hardware connector and a mobile phone connected via bluetooth.

As mentioned above, the On-Board-Diagnostic interface allows for interaction with a variety of ECUs and to obtain valuable data for an overall insight into the current state of a vehicle. 
Especially telecommunications service providers such as Vodafone \cite{vodafone}, Telekom \cite{telekom} or Telefonica \cite{telefonica} offer OBD-II connectors in combination with a mobile data connection. 
Note, this also broadens the attack surface as it could provide remote access to the CAN (Controller Area Network) bus, transforming a perceived internal attack surface (needing physical access to the OBD-II port) into an external threat \cite{7030108}, \cite{7952095}. 
But apart from that, these service providers gain insights into an enormous amount of highly private data about the drivers' everyday life \cite{website:newscardata}. 
Among other things, this includes the driving style, routes and accurate driving times. Manufacturers and insurance companies as well as service providers from the business sector are already using the information provided as a basis for monetarizing the individual journeys of users. 
An example would be the reduction of car insurance if you have a particularly restrained and safe driving style. 
This monetarization can also work in the opposite direction. 
This means that drivers who drive very aggressively, for example, are punished with higher rates. 


In summary, OBD-II communication creates two new threat categories: 
\begin{enumerate}
\item Most obvious is the problem of  malicious inbound flow from an adversarial dongle to the car
\item less obvious --judging from the limited amount of research so far-- is the problem of personal data leakage in outbound flow from the car to the dongle
\end{enumerate}
As there is no encryption or authentication on the CAN bus by default \cite{Groza2018SecuritySF}, not only inbound threats are very real and far from difficult to implement \cite{Groza2018SecuritySF}, but also the ability to tamper with data on its way from the car to the dongle is not prohibited due to the --by default-- missing data origin-authentication.

\subsection{Goals and Contributions}
All this raises the question of whether it is possible to develop an effective and modular firewall-like filtering approach for data flows in both directions via the given interface. 
We therefore propose a rule-based approach suitable for the OBD-II interface and all affected protocols to protect the driver and his car from possible malicious dongles by modifying or rejecting data flowing in both directions. 

The idea of something like a firewall in a car was proposed by Rivzi et al. where they showed a distributed firewall approach and put an inbound filter on each ECU in a vehicle \cite{rizvi:firewall}.
However, our approach is specifically designed to manipulate traffic, especially the content of traffic from the car to the dongle. 
To the best knowledge of the authors this has not been done before and our approach  still allows full or reduced operation of the dongle, unlike current commercially available blockers, which only offer an all-blocked or all-allowed approach.
Let us stress that being able to manipulate traffic on OBD-II is an attack vector on its own, highlighting an absence of crucial security mechanisms for this communication channel. 
We exemplify the attack by building a Man-in-the-OBD-II interface
to highlight 
how the data flow 
accessible
via the OBD-II interface can be modified or restricted. One of the hurdles to be considered is the provision of the firewall for the end user. The objective is to make it as easy as possible for users to enter the firewall without any major entry hurdles. Therefore, in Section \ref{chap_spefification:section:policymanagement} we introduce a configurable policy language in the well-known JSON format, which eliminates this problem.


Our modular design works on the standardised OBD-II stack and we have chosen to implement both inbound filtering and outbound filtering for Controller Area Network protocol messages. CAN 2.0 specification was published in 1991 by Robert Bosch GmbH \cite{CANBOSCH} and standardized by ISO 11898-1 \cite{ISO11898-1} first in 2003, and is one of the most common used protocols for vehicular information exchange through dongles today. 
However, since there are many different communication mechanisms and protocols in the OBD-II stack, our general approach abstracts from the protocol itself.
For brevity we focus in this work on CAN message filtering and rewriting in both directions, which can be seen as one module in our general OBD-II man-in-the-middle firewall concept. This makes it possible to easily add other protocols to our basic system with a functioning implementation for CAN. It means that our protocol-agnostic approach also enables manufacturer-specific solutions. The only hurdle that remains for anyone who wants to use the firewall is to configure the correct rules using our rule language. For this, the user must be able to understand the relevant important data, e.g. the message format within CAN, and know exactly which commands are sent. In the case of CAN, however, there is already a lot of current research that helps to reverse engineer messages (\cite{libreCAN}, \cite{automatic_reverse_CAN} and \cite{READ_reverse}).
%
%
\subsection{Outline}
In this work 
we 
focus on two aspects: 
\begin{enumerate}[I:]
\item Identify the threats a security layer in between the OBD-II dongle and the car would solve
\item Implement such a security layer as a Man-in-the-Middle for the OBD-II and show how it can fool dongles to protect the car driver's privacy
\end{enumerate}
%
%

We present an overview of related work in Section 2.
In Section 3 we facilitate the standard threat modeling tools DREAD and STRIDE to  analyze systematically the possible threats that can occur and, in the best case, can all be avoided by implementing our approach. 
We 
highlight 
the most important threats by means of this procedure.
In Section 4 we present the architecture of our solution that abstracts the dongle from the actual vehicle network and shields the car as best as possible. 
%
%
In Section 5 we discuss briefly the implementation of our approach 
for the case of CAN messages. 
We 
provide 
a rule-based policy language. 
By means of different options and types within the 
rules it 
is possible to set the firewall filters
for inbound and outbound data flows. 
%
We evaluate the impact on the identified threats and on selected dongles in Section 6.
Finally, we conclude in Section 7.
%
%

\section{Related Work}
\label{related_work}
Work has been done on the vulnerabilities added by the dongles itself: 
The idea of analysing dongles' behaviour
was discussed by
Wen et al. \cite{247700}; they provide a comprehensive vulnerability analysis of 77 OBD-II dongles. In the paper the authors propose an automated tool called DongleScope to perform an analysis and to test the dongles. 
In the paper published by Yadav et al. \cite{yadav2016} the authors give an overview of various security vulnerabilities and points of entry for malicious entities in vehicular systems. 

The trove of information that can be gained from a car
is shown in several works 
(\cite{website:newscardata}, \cite{8252037}, \cite{8300417} and \cite{8492706}); 
they 
show how to monitor automobiles, predict the condition of the internal hardware, detect driving behaviors and discover different anomalies.

In the remainder of the related work section we grouped the works by their view point on the information flow.

\subsection{General vehicular security concepts describe the threat of unwanted information flow}
In the paper by Bernardini et al. \cite{BERNARDINI201713}, eight security requirements and five safety requirements for vehicle systems are defined and explained. They also describe how existing systems and solutions such as the AUTOSAR architecture, LIN, FlexRay, MOST and Ethernet/BroadR-Reach are aligned and can be used to fulfil these requirements. Furthermore, the authors explain in detail which safety concepts are to be pursued in vehicle systems and which possible problems or limitations may arise. 

Hoppe et al. \cite{10.1007hoppe} show that it is necessary to examine and possibly modify already existing security vehicle systems. Among other things, the authors show that the intrusion detection system does not fully protect cars from intruders. Even though this system is one of the newer ones in vehicle safety, according to Hoppe et al. some improvements need to be made to minimise the risk of attacks. This shows us that even current security concepts are often not fully developed and cannot offer complete protection. Therefore, our approach is to enable a security layer as simply as possible according to the plug and play concept. 

Studnia et al. \cite{studnia:hal-00848234} have looked at fundamental problems related to car security. Among other things, they found that the computing power of a car is very limited and this can lead to problems when using strong cryptography within certain protocols. In addition, they figured out that car manufacturers must validate the software running on an ECU embedded within a vehicle and test it on a periodic basis to guarantee its integrity. An entire vehicle can become vulnerable if bugs remain in the vehicle system. These effects are of course reflected in the severity of the respective bug. In the event that a security flaw is utilised, it can require anywhere from several months to years for a patch to be installed for all of the specific cars that were already on the road. This implies that it is an extremely important task to prevent malicious code from entering the vehicle in the first place. Therefore, our solution is to safeguard the OBD-II interface and thus exclude possible attackers.
%
\subsection{Filtering of inbound traffic towards the car's ECUs exists}
\label{sub:filtering_inbound_traffic}
Wolf et al. \cite{Wolf2004SecurityIA} examined the prevailing architecture as well as the threats that are prevalent in contemporary vehicles. They discovered that the gateways built into the automotive network require the use of powerful firewalls. In addition to this, they stated that the firewall implemented in the gateways also need to possess rules that control access on the basis of the security relevance of the particular network.

While filtering approaches or firewall concepts for networks inside the vehicle do exist and are nothing completely unknown, the range of available research is very limited, especially compared to works on inter-vehicle networks like VANETS. Even less information is available about  existing solutions for cars being deployed. 
For example NXP describes the need to protect the car's networked devices from unwanted outside traffic by a gateway for 
''filtering inbound and outbound network traffic based on rules, disallowing data transfers from unauthorized sources.``\cite{nxp_gateway_whitepaper}.
NXP further states that a more fine-granular approach ``[...] may include context-aware filtering`` \cite{nxp_gateway_whitepaper}. 
But often the exact mechanisms used as well as the security functions in real vehicles are not publicly published. 
Another manufacturer's solution is the ``Central Gateway'' for central in-vehicle communication from Bosch, which lists a firewall and an intrusion detection system on its product page \cite{bosch_gateway}. However, neither the info PDFs nor the actual page list more precise details. Even when specifically asked at the responsible department, we were unable to get any further information about the security features mentioned.
The company  Karamba Security \cite{karamba_security} in 2016 released a security architecture that acts as a gateway between a car's access points and critical networks/modules. Karamba calls it ECU Endpoint Security: Dropper Detection and Malware Prevention. To define factory policies, the developers had the idea of having a system embedded directly in the firmware. This is to prevent malicious code from infiltrating the system. Each ECU specifies its own policy and generates a so-called whitelist of permitted programm binaries, processes, scripts and network behaviour. 

In academic literature Rizvi et al. presents this as a distributed approach for a firewall system in automobile networks \cite{rizvi:firewall}: Their system is focused to let only authorised packets reach an internal device using a Hybrid Security System (HSS) that uses many individual firewalls located in front of each module and at each electronic unit. 

\subsection{Commercial available approaches towards filtering and securing the OBD-II Interface}
Practical OBD devices which do offer such inbound filtering or just blocking all access are available on the market. 
The most critical feature touted as blockable by most entry filters is the use of ``key duplicators and an accessible OBD-II socket`` that would allow car thieves to generate new access codes thus obsoleting the original keys. This creates a safety barrier between the external devices and the data bus protecting vehicle functions against unauthorized access and manipulation. Using key duplicators and an accessible OBD-II socket, a car thief can easily generate new access codes, outflanking in a few moves the existing car alarm system. As can be clearly seen in Table \ref{tab:overview_commercial_products}, almost all approaches get delivered with only two modes supplied: always-deny (Off) or always-allow (On). 
Their focus is on preventing malicious senders' packets from reaching important devices in the vehicle by turning the CAN bus access off.
Of course that would also completely block data that might travel in the other direction, but this privacy impact is not advertised.  
Furthermore, with the existing approaches, the use of an OBD-II dongle is not possible when activated, as absolutely no data is available for processing. Our approach closes this gap.

\begin{table}[]
\label{tab:overview_commercial_products}
\caption{Commercially available products' capabilities compared to our Man-in-the-OBD-II approach}
	\centering
	\resizebox{\textwidth}{!}{%
\begin{tabular}{|l|c|c|c|c|}
\hline
\multicolumn{1}{|c|}{\textbf{Product}} & \textbf{\begin{tabular}[c]{@{}c@{}}Operation \\ Types\end{tabular}}        & \textbf{Modes}                                                                             & \textbf{Filtering}       & \textbf{Method} \\ \hline
Diagnostic BOX - OBD Blocker \cite{diagnostic_box_obd_blocker}          & None                                                                       & On/Off                                                                                     & None                       & MiM             \\ \hline
Ampire CAN-BUS Firewall \cite{ampire_can_firewall}               & None                                                                       & On/Off                                                                                     & None                       & MiM             \\ \hline
Ampire OBD-Firewall \cite{ampire_obd_firewall}                    & None                                                                       & On/Off                                                                                     & None                       & MiM             \\ \hline
Paser Firewall OBD2 \cite{paser_firewall_obd}                    & None                                                                       & On/Off                                                                                     & None                       & MiM             \\ \hline
Electronic anti theft OBD plug \cite{electronic_anti_theft_obd}         & None                                                                       & On/Off                                                                                     & None                       & MiM             \\ \hline
AutoCYB \cite{auto_cyb_lock}                               & None                                                                       & \begin{tabular}[c]{@{}c@{}}Mounted/\\ Unmounted\end{tabular}                               & None                       & Lock            \\ \hline
CAN Hacker Diagnostic Firewall \cite{can_hacker_firewall}        & Unknown                                                                    & Unknown                                                                                    & SIDs/PIDs                  & Unknown         \\ \hline \hline
\textit{Man-in-the-OBD-II}             & \textit{\begin{tabular}[c]{@{}c@{}}reject, limit, \\ replace\end{tabular}} & \textit{\begin{tabular}[c]{@{}c@{}}delay, pub\_once,\\ id\_range, val\_range\end{tabular}} & \textit{Individual} & \textit{MiM}    \\ \hline
\end{tabular}
}
\end{table}


\section{Threat Modelling for OBD-II} 
\label{sec:threat_identification}
We conducted a threat analysis for a commercial passenger car with an OBD-II dongle. 
Threats were categorized according to the STRIDE method \cite{STRIDE}. 
We then use the DREAD method \cite{DREAD} to rate the seriousness of those threats. 
\subsection{Threats following STRIDE}
For brevity we limit this to the most relevant 
eight threats, which are 
used 
to draw a 
conclusion about the effectiveness or usability of our approach 
in Section \ref{threat_impacts_with_firewall}.

\paragraph{\textbf{$T_\alpha$}}
\textbf{Malicious Device plugs directly into OBD-II}

This threat is virtually impossible to prevent. As soon as the attacker has physical access to the interface, he can, for example, bypass all upstream hardware security modules (such as our firewall approach) or simply plug them off. $T_\alpha$ could only be effectively prevented by additional physical security mechanisms. One possibility would be to physically separate the hardware security modules from the accessible OBD-II interface in such a way that it is no longer possible for an attacker to either:
\begin{itemize}
	\item Separate the module from the vehicle, or
	\item the use of the OBD-II interface becomes unusable as soon as the module is disconnected.
\end{itemize}
However, this would require some modification of the OBD-II standard and is not really feasible. 

\paragraph{\textbf{$T_\beta$}}
\textbf{Attacker compromises a running service on dongle}



The threat class $T_\beta$ is a class which includes threats that may or may not occur frequently, depending on the security standards applied in the development of the software running inside the dongle. 
The attack surface is very large due to the use of cloud services, an internet interface (e.g. by means of a SIM module in the dongle) or a Wi-Fi interface. The very fact that the service communicates with a cloud service via an API means that in addition to the service inside the dongle also the cloud services' endpoints must be regularly maintained and also updated. If this is not done, the risk of a possible attack increases further. 
In the following, there are two subcategories in the $T_\beta$ class, namely:

\begin{enumerate}[I:]
	\item Dongle refers to hardware where our firewall runs on
	\item Dongle refers to the third-party hardware that plugs into our firewall 
\end{enumerate}
\paragraph{\textbf{$T_\gamma$}}
\textbf{Attacker can send arbitrary CAN commands}

This threat is enabled by either $T_\alpha$ or $T_\beta$. The attacker is thus able to control, for example, displays in the instrument cluster or the opening and closing of the electric windows. Generally speaking, it is possible for him to contact all ECUs that communicate via CAN and can be reached via the OBD-II interface. This enables a wide range of attack vectors. 

\paragraph{\textbf{$T_\delta$}}
\textbf{Attacker can read communication on CAN-Bus}

No direct physical damage can be caused by this threat, as it is only a matter of reading access to the CAN bus. Nevertheless, the privacy of the respective user can be violated. It is possible to listen to all communication on the CAN bus that is accessible via OBD-II. By analysing the traffic, a variety of conclusions can be drawn.  For example, statements can be made about the individual driving behaviour of each driver in road traffic. 

\paragraph{\textbf{$T_\epsilon$}}
\textbf{Damaging ECUs by executing specific commands}

In order to damage individual ECUs by means of specific messages, the attacker needs a profound understanding of the respective vehicle system as well as the ECU to be damaged. In most cases, such an attack is not possible without extensive prior testing and analysis of the hardware. Of course, one can actively try to damage the hardware by sending harmful and unwanted CAN messages. However, a try-and-error approach offers little chance of success.

\paragraph{\textbf{$T_\zeta$}}
\textbf{Person endangerment by deactivating safety functions}

This threat is a very dangerous one, as it may actively cause physical harm to people. On the one hand, an attacker could succeed in deactivating critical safety systems such as the airbag, the ABS (anti-lock braking system) or the ESP (electronic stability program). Thus, in the event of a driving manoeuvre where the respective system is required, a possible accident would be the result. Furthermore, an attacker could, for example, display the wrong speed. As an example, he could simply display a much lower speed than is actually being driven at the moment. Thus, the driver could be harmed if he is at that time in a certain road passage where maintaining a certain minimum speed is critical to safety.

\paragraph{\textbf{$T_\eta$}}
\textbf{Permanent infiltration of the vehicle system by uploading malware}

One 
goal
during an attack or after a successful intrusion into a system is to make the access persistent. 
$T_\eta$ describes the maintenance of unrestricted access even after possible malicious OBD-II devices have been removed. This requires in-depth knowledge of the ECUs available in the target vehicle. Possible targets are especially ECUs that have a memory module or dedicated firmware that can be overwritten.
\begin{figure}
\vspace{-1em}
\begin{floatrow}
\vspace{-1em}
\ffigbox{%
  \hskip-0.2cm\includegraphics[width=0.5\textwidth]{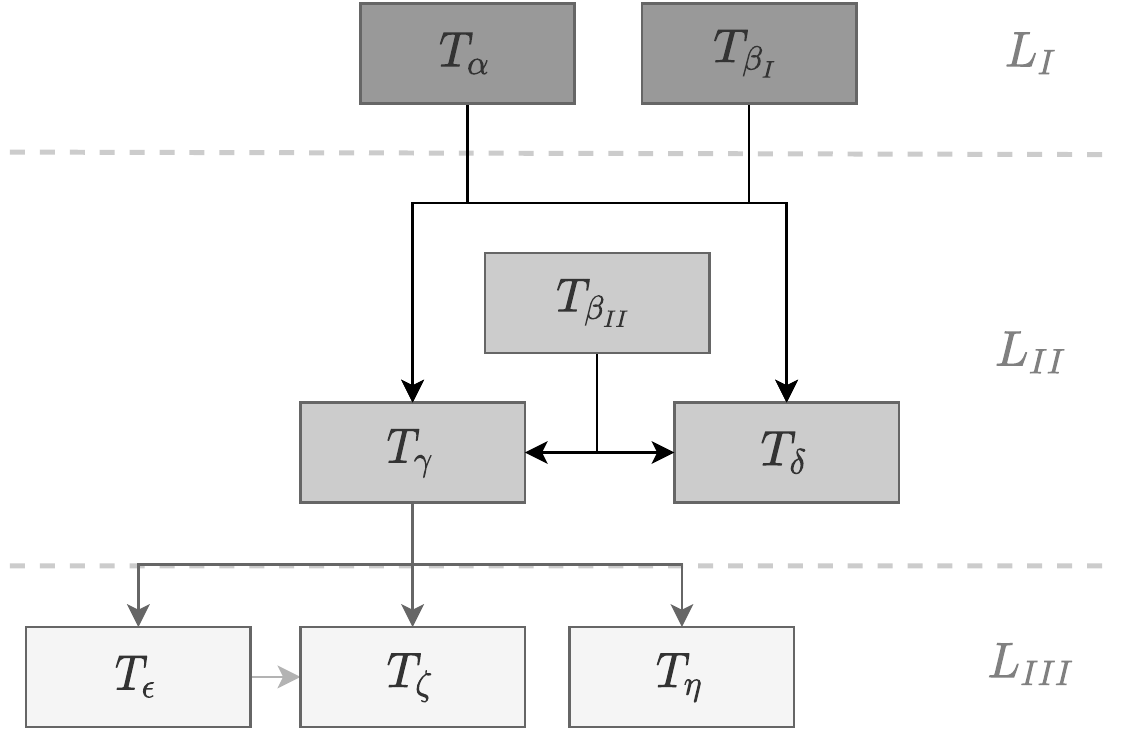}
  
}{%
  \caption{Attack tree of specified threats}
  \label{fig:threats_attack_tree}
}
\capbtabbox{%
  \hskip-0.6cm\begin{tabular}{|c|c|c|c|c|c|c|c|c|}
		\hline
		\multicolumn{1}{|l|}{}             & $T_\alpha$ & $T_{\beta_I}$ & $T_{\beta_{II}}$ & $T_\gamma$ & $T_\delta$ & $T_\epsilon$ & $T_\zeta$ & $T_\eta$ \\ \hline
		\textbf{D}                         & 3            & 3 & 3        & 3            & 2            & 3             & 3           & 3 \\ \hline
		\textbf{R}                         & 3            & 2 & 2          & 2           & 2            & 2              & 2           & 1 \\ \hline
		\textbf{E}                         & 3            & 2 & 2          & 2           & 2            & 1              & 2           & 1 \\ \hline
		\textbf{A}                         & 3            & 3 & 3         & 3           & 3            & 3              & 3           & 3 \\ \hline
		\textbf{D}                         & 3            & 2 & 2        & 3            & 3            & 2              & 3           & 2 \\ \hline
		\multicolumn{1}{|l|}{DREAD Risk} & 3.0            & 2.4 & 2.4         & 2.6            & 2.4             & 2.2               & 2.6           & 2.0 \\ \hline
	\end{tabular}
	
}{%
  \caption{DREAD rating of selected threats}
  \label{tab:obd_dread_table}
}
\vspace{-1cm}
\end{floatrow}
\end{figure}
\subsection{Results of OBD-II Threat Modeling}
Based on the seven overall threat classes defined and explained above, 
we 
used the DREAD rating model \cite{DREAD} to 
roughly classify how big or serious the individual threats are. 
The individual values for each threat 
can be seen in Table \ref{tab:obd_dread_table}. 
Levels range from low (1) to high (3) risk. 
When comparing the individual result values, it becomes clear that $T_\alpha$ is the highest rated threat with a risk rating of 3.0. The main reason for this is that $T_\alpha$ is a physical threat. With physical threats, the possibilities of an attacker are always greater than, for example,  in the case of a remote only attack. The lowest rated threat is $T_\eta$ with a risk rating of 2.0. This is because this type of attack requires a tremendous amount of knowledge and skill to execute. Usually a lot of research and testing needs to be done on the real physical devices/ECUs to even find vulnerabilities that make it possible to carry out the attack. It is also interesting that, with the exception of threat $T_\delta$, the damage for all other threats is always rated at the maximum of three (\textit{high}). The reason for this lies in the effect of the respective threats. $T_\delta$ describes the possibility to read messages of the CAN bus. This means that a possible attack can also be described as passive, since it never actively changes anything on the BUS or writes anything to it. Nevertheless, sensitive data (e.g. regarding the driver's privacy) can be collected by reading and possibly decoding some messages. Therefore, $T_\delta$ still has a rating of two, which means medium in terms of the DREAD rating system.

Since some threats enable other threats, we have created an attack tree of the seven selected threat classes. An analysis via attack trees provides a graphical, easy-to-understand modelling of threats. It helps us to classify the different attack possibilities of the OBD-II interface more precisely and to develop possible countermeasures to prevent such attacks. In Figure \ref{fig:threats_attack_tree} you can see this tree, which can also be divided into three hierarchies named $L_{I}$, $L_{II}$ and $L_{III}$.

The first level is the basic threat hierarchy. This means that threats from higher levels are always based on threats from the basic level. The threats $T_\alpha$ and $T_{\beta_I}$ are therefore needed to realise attacks with the threats from the layers below. Therefore, we also define these two threats as entry threats. In the best case, all entry threats can be eliminated or prevented or mitigated so that all further threats become obsolete or not quite as severe. The threats in our last level $L_{III}$ are the most specific in terms of the knowledge required or the techniques used. 

\section{Architecture of the Man-in-the-OBD}

In this section we 
describe the architecture and the individual abstract 
components, which we will then implement in Chapter \ref{chap:implementation_and_evaluation}. 
For brevity we only briefly explain the basic ideas of the respective components theoretically and show why we choose exactly this approach. 

\subsection{Producer/Consumer Scheme}
\label{sub:producer_consumer_scheme}
The producer-consumer problem (also known as the bounded buffer problem) is a classic example of a multiprocess synchronisation issue, the first version of which dates back to Edsger W. Dijkstra in 1965. Nevertheless, there are now promising approaches in software development to efficiently eliminate this problem \cite{10.1145/3335772.3335782}. We have decided on such an approach (more details can be found in Section \ref{sub:producer_consumer_solution}). A filtering approach is best realised with a buffer in which incoming messages are accumulated. Afterwards, they can be processed one after the other, depending on the queue. This model is ideal if you want to be as unrestricted as possible in the processing phase. Depending on the respective computing power, several producers or consumers can be started. In this way, load peaks can be easily absorbed. The modular approach can also be applied by means of differently implemented producers.

\subsection{Modular Approach for Protocol Bindings}
Since the producer/consumer scheme allows us to easily create several differently implemented producers, a uniform interface must be defined. This interface ensures that the responsible consumer can correctly process and forward the incoming messages. By means of this approach it is possible to support incoming messages of all protocols. With this method, a high-performance and efficient filtering is possible.

\subsection{CAN-Bus Binding}
In order to support the CAN protocol for our implementation, a connector is needed to receive messages as well as to be able to send filtered or processed messages again. For this purpose, already widely used libraries (e.g. \cite{cantools}, \cite{porcelain}) as well as the common syntax for coding and decoding are used. Furthermore, it is desirable if the binding understands the so-called DBC format. DBC stands for CAN Database and is a proprietary format that describes the data structure over a CAN bus. A CAN~DBC file is a text file that provides all the necessary information for decoding CAN bus raw data into physical values. If a DBC file is available for the respective manufacturer, the user can easily decode the CAN data streams and thus analyse them in order to manipulate or block them as desired in our firewall approach. This kind of provision can ensure a possibility to support as many manufacturers as possible. Analogous to this binding, a binding for ISO~9141 or SAE~J1850 could also be written and integrated into our pipeline as a producer in order to support these protocols.

\subsection{Processing Pipeline}
\begin{figure}
    \centering
	\includegraphics[width=0.95\textwidth]{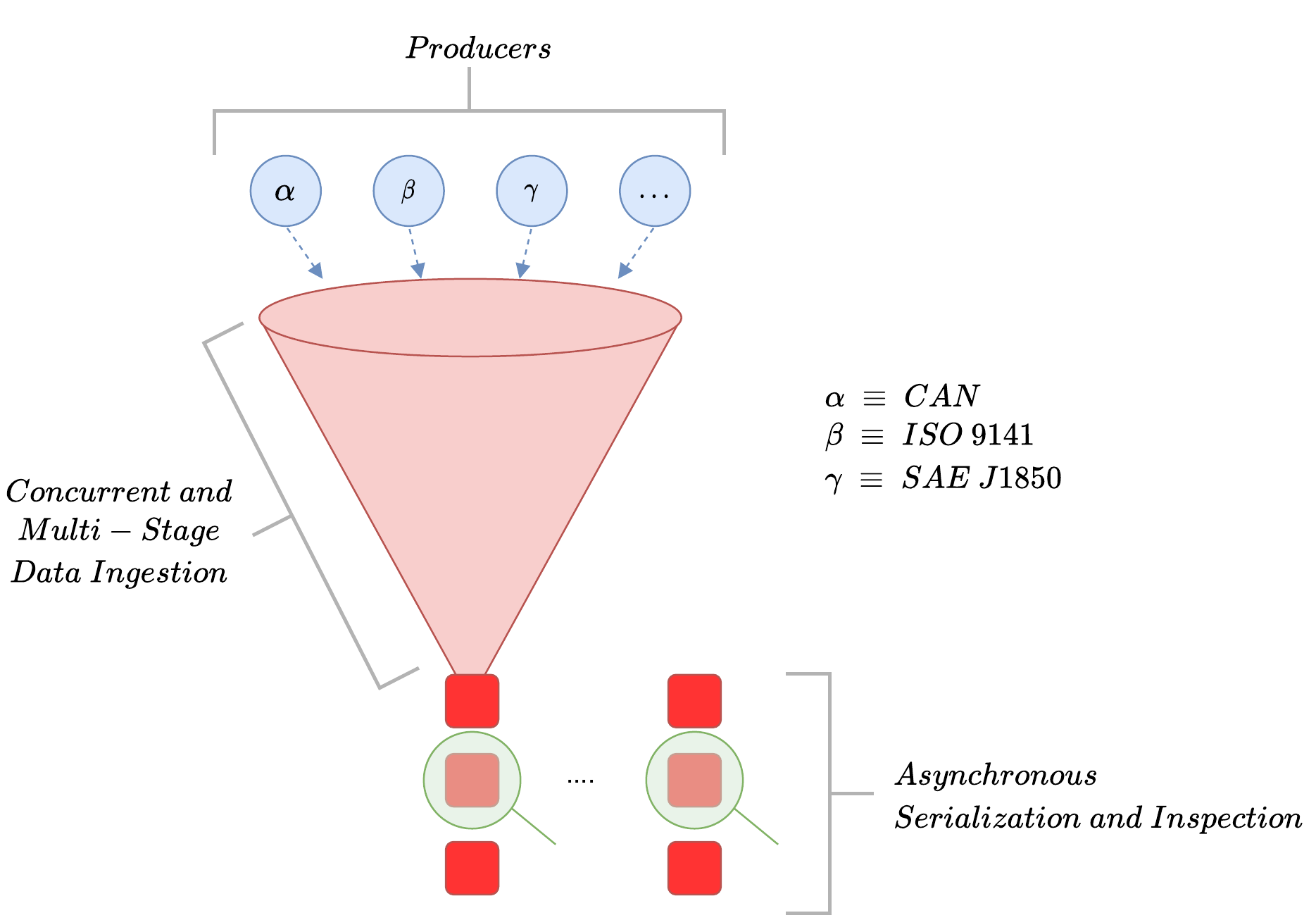}
	\caption{Simplified representation of the processing pipeline}
	\label{fig:software_architecture_data_generation}
\end{figure}

A concurrent and multi-level data input and data processing pipeline is to be realised for the processing pipeline. It must also be possible to efficiently consume different sources, the so-called producers. It would also be desirable if there was the possibility to configure the processing pipeline with regard to the resources to be used. More precisely, the number of processes for producer and consumer as well as the concurrency and the batch size to be used. In Figure \ref{fig:software_architecture_data_generation} you can see an schematic overview of the pipeline. The blue circles in the figure symbolise the producers. As already mentioned, it will be possible to develop a producer for each protocol. So in the future there may be a producer for CAN, one for ISO~9141 and so on. There is a uniform interface to adhere to. The producers then send their messages to the concurrent and multi-stage data ingestion service. There, the incoming messages should be analysed and then asynchronously serialised and inspected. Here, serialisation refers to the application of the active rules and not to the quantity of messages.

\vspace{-2.5em}
\subsection{Serialization}
After the incoming messages have been serialised and bundled into batches, the messages are to be checked for the active rules as efficiently as possible. Since the behaviours of a rule can be sorted according to their strictness in the restriction, the strictest behaviours should currently apply. This allows the individual behaviours to carry out the checks in parallel. 
\vspace{-1em}
\subsection{Data Storage}
It should be possible to record the data as well. In the best case, data should be stored in a database in a uniform, reusable format. This ensures that the logged CAN messages can be easily searched or filtered for various purposes. Since the amount of CAN messages can be immense, there should be an option to deactivate or activate the permanent logging.

\vspace{-1em}
\subsection{Policy management}
\label{chap_spefification:section:policymanagement}
Policy management systems can be implemented either as specialised hardware or as software on general-purpose operating systems. However, the underlying idea is always the same. There is a set of defined rules that determine which packets the separated network can receive and how those packets are modified if necessary \cite{8406593}. The best known firewall tools under Linux and Unix are \textit{iptables} and \textit{ipfw}. However, since these are far too extensive and complex for our current needs, we have opted for a simple implementation of our own. Other well-known policy languages such as RPSL \cite{rfc7909}, SRL \cite{rfc2723}, PAX \cite{nossik-pax-pdl-00}, PFDL \cite{ietf-policy-framework-pfdl-00} may also not be suitable for our application and often the entry hurdle would be much higher than with our simplified policies in the form of the widely known JSON format.  These facts are the main reasons why we do not currently use a specific rule framework or rule engine. As briefly touched on, our policies are managed using configurations specified in JSON format. The format is a simple one that is adapted to the current use case, but can be extended in a modular way. The current overall structure can be seen in Tables \ref{tab:policy_properties_list} and \ref{tab:policy_property_behaviour}. In any case, a kind of version check must be carried out at implementation stage for the respective rules to be applied. On the other hand, an extension is almost impossible or backwards compatibility cannot be guaranteed. Table \ref{tab:policy_properties_list} shows the overall wrapper structure for a rule definition. This contains general information such as a description, the protocol type to be filtered and the version of the policy language currently in use. Table \ref{tab:policy_property_behaviour} describes the structure of a so-called behaviour. A rule can theoretically have as many behaviours capsules as desired. Here, each currently available behaviour type (namely reject, limit and replace) is applied once for demonstration purposes.

\subsection{Rule enforcement}
The enforcement of rules as well as individual behaviours is based on an assessment of importance. This means that more important behaviours and rules outweigh less important ones. For this purpose, there is a special type rating, which is specified using the type property of the behaviours. As in Section 4.5, this has the great advantage that the individual behaviours can be executed in parallel after the initial filtering of the strictest rating and thus best fit our scalable approach.

\newpage
\begin{table}[H]
	\centering
	\caption{List of basic properties with their associated functionality}
		\begin{tabular}{|p{0.13\textwidth}|p{0.18\textwidth}|p{0.69\textwidth}|}
			\hline
			\textbf{Property}    & \textbf{Type}   & \textbf{Description} \\ \hline \hline
			name        & \textit{$<$String$>$}        & Is just a simple naming of the individual rules for better distinction. The name does not have to be unique.              \\ \hline
			description & \textit{$<$String$>$}        & Briefly describes the created rule in a few words.             \\ \hline
			version & \textit{$<$String$>$}        	  & The version number specifies the version of the properties to be used.              \\ \hline
			protocol    & \textit{$<$Protocol-Type$>$}       & Declares the protocol type to be used for the respective rule. Currently there is only \textit{$<CAN>$} as a declarable type.       \\ \hline
			behaviours  & \textit{$[<$Behaviour$>]$}        & The behaviour field defines a list of all actions to be performed later during the execution of each rule.              \\ \hline
		\end{tabular}
	\label{tab:policy_properties_list}
\end{table}
\vspace{-3em}
\begin{table}[H]
	\centering
	\caption{List of properties for a single behaviour inside a policy\\}
		\begin{tabular}{|p{0.12\textwidth}|p{0.14\textwidth}|p{0.72\textwidth}|}
			\hline
			\textbf{Property}    & \textbf{Type}   & \textbf{Description} \\ \hline \hline
			type        & \textit{$<$String$>$}        & Currently, three different behaviour types are supported: \begin{itemize}
				\item \textit{reject} - Ignores all messages with the defined identifier and associated value
				\item \textit{limit} - Limits all accruing values of a message from the defined identifier by means of a predefined value
				\item \textit{replace} - Always exchanges all message values of a given identifier with the given value
						\end{itemize}             \\ \hline
			identifier & \textit{$<$String$>$}        & Defines the identifier of the CAN message present on the bus           \\ \hline
			value    & \textit{$<$String$>$}       & Determines the data payload to be used for the respective set type            \\ \hline
			\multicolumn{3}{|c|}{\textit{The following properties are optional and do not have to be set}} \\ \hline
			delay    & \textit{$<$Integer$>$}       & If the delay property is set, all messages that fall below the specified behaviour will be delayed. The value is given as an integer value and defines the delay time in milliseconds.             \\ \hline
			pub\_once    & \textit{$<$Boolean $>$}       & Allows messages in the scope of the behaviour to be allowed only once per system start. Once the message has been read once, it is whitelisted and then not forwarded. By default, the value is set to \textit{false}.              \\ \hline
			id\_range    & \textit{$<$String$>$}       & By means of the identifier range, the behaviour value range to be enforced can be extended.              \\ \hline
			val\_range    & \textit{$<$String$>$}       & Allows messages in the scope of the behaviour to be allowed only once per system start. Once the message has been read once, it is whitelisted and then not forwarded. By default, the value is set to \textit{false}.              \\ \hline
		\end{tabular}
	\label{tab:policy_property_behaviour}
\end{table}

\newpage
\section{Implementation}
\label{chap:implementation_and_evaluation}
Next we describe all the components that run our developed approach in the background. This includes the processing of incoming messages, the used producer \& consumer approach, the storage of individual messages as well as our filter module.

\subsection{Producer/Consumer Solution}
\label{sub:producer_consumer_solution}
As already described in Section \ref{sub:producer_consumer_scheme}, our approach should benefit from the so-called producer consumer construct and thus make a modular approach more feasible. This is one of the disciplines in which Elixir can demonstrate all its abilities and advantages in the best possible way. To build our solution, we use the library called \textit{Broadway} \cite{broadway}. 

\begin{figure}
	\centering
	\includegraphics[width=0.7\textwidth]{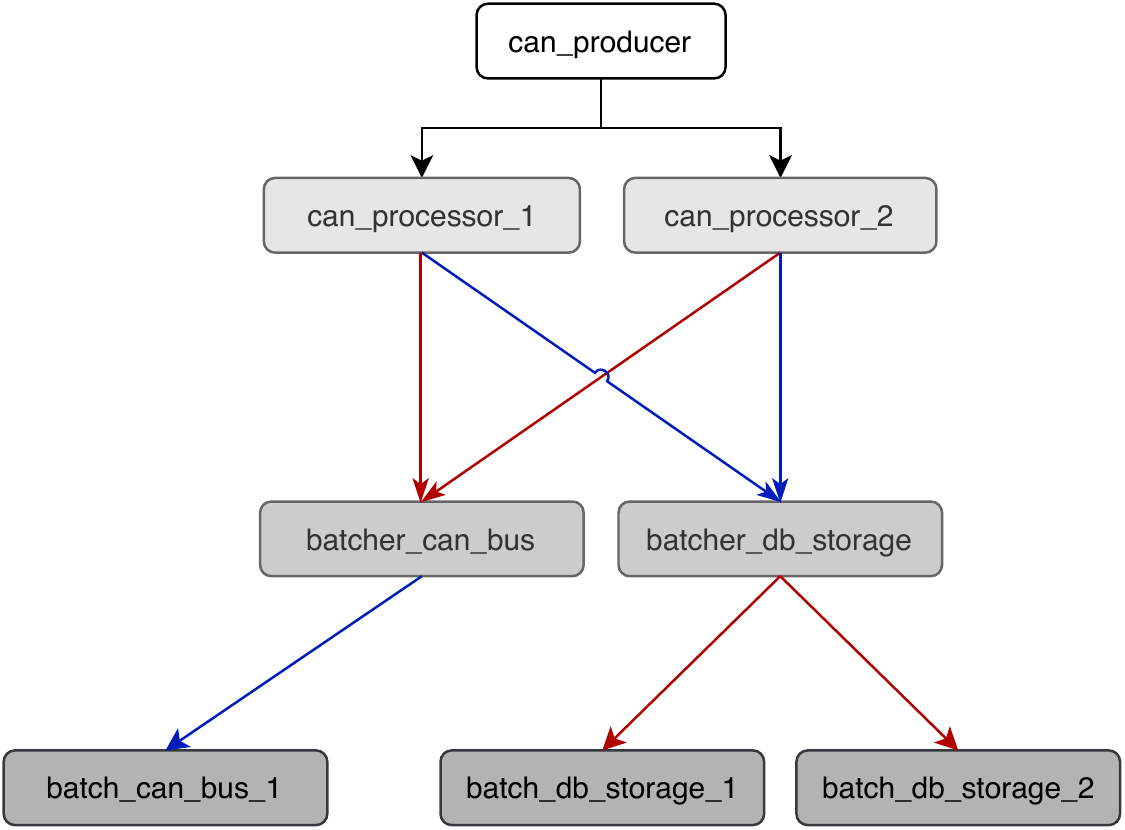}
	\caption{Representation of producer, processor and batcher pipeline}
	\label{fig:broadway_pipeline}
\end{figure}

The library can be used to create concurrent, multi-stage data input and data processing pipelines. 
It is also possible to implement your own producers, which is perfect for our use case. Depending on the use case, it can make a lot of sense to summarise the processed messages as a so-called batch before the actual publication. 
While we don't need the batchers to communicate with an extraneous API, it allows us to process the storing of CAN messages in an encapsulated way and thus have no runtime loss for publishing already filtered CAN messages. This allows for increased throughput and consequently improved overall performance of our pipeline. Batches are simply defined via the configuration option. The configuration is of course adaptive and can be extended very easily if required. A schematic representation of how our current testing pipeline looks is shown in Figure \ref{fig:broadway_pipeline}.

\vspace{-2em}
\subsection{Storing of CAN-Messages}

The storage of CAN messages can be set via the web user interface. By default, CAN messages are not saved. However, if the option is activated, all filtered messages are stored in a PostgreSQL database in the \texttt{batcher\_db\_storage}. Since the storage is carried out in one of the batcher processes, all storage operations can be carried out completely independently of the runtime of the actual filter pipeline. PostgreSQL was chosen because the \textit{Phoenix framework} uses it as the default database. The advantage of this, however, is that the adapter responsible for the connection to the database can simply be exchanged. This makes it easy to switch to a Time Series Database (TSDB) such as InfluxDB \cite{influx} or Riak \cite{riak}. TSDBs are suitable because CAN messages are time-distributed data. Since the storage and further processing of past CAN bus data is not relevant for our work, the use of PostgreSQL is sufficient for our purposes.

In addition to the raw data of the identifier and the payload which are stored using the \texttt{byte} data type, we also store the encoded data as \texttt{varchar(255)}. This has the advantage that we can display the data in a simple and user-friendly way in the front-end without having to decode a large number of data sets each time.

\subsection{Pipeline Benchmarks}

In Table \ref{tab:benchmark_results} you can see the measurement results for the runs of our pipeline with different numbers of behaviours. 
Our firewall runs on a Raspberry Pi 4 with 8 GB RAM.
The measurements generally show that our filtering pipeline only requires a small amount of computing time and therefore has no impact on the usability of the dongles. The average runtime of the pipeline increases with the increase in the number of behaviours defined in a rule. In general, the additional execution time has no effect on the correct functioning either way, as the dongles are inherently time uncritical (more on this in Section \ref{conclusion}).
\vspace{-2em}
\begin{table}[H]
	\centering
	\resizebox{\textwidth}{!}{%
		\begin{tabular}{|l|l|l|l|l|l|l|l|}
			\hline
			\textbf{Measurement}    & \textbf{Iterations per Second} & \textbf{Average} & \textbf{Deviation} & \textbf{Median} & \textbf{Minimum} & \textbf{Maximum} & \textbf{Sample Size}                   \\ \hline
			3 behaviours & 243.87 & 4.10 ms & $\pm$9.71\% & 4.08 ms & 3.05 ms & 6.79 ms & 1218 \\ \hline
			30 behaviours & 225.89 & 4.43 ms & $\pm$11.89\% & 4.36 ms & 3.29 ms & 10.23 ms & 1218 \\ \hline
			300 behaviours & 164.58 & 6.08 ms & $\pm$11.90\% & 6.03 ms & 4.45 ms & 11.07 ms & 823 \\ \hline
			3000 behaviours & 83.22 & 12.02 ms & $\pm$7.44\% & 12.02 ms & 9.55 ms & 14.68 ms & 416 \\ \hline
			30000 behaviours & 16.74 & 59.72 ms & $\pm$7.09\% & 59.10 ms & 51.62 ms & 69.53 ms & 84 \\ \hline
		\end{tabular}%
	}
	\caption{Results of the run time comparison of the single measurements}
	\label{tab:benchmark_results}
\end{table}

\vspace{-2em}
\section{Evaluation of Impact on Threats and Existing Dongles}
	To evaluate we have tested our approach with real software currently available on the market; we report the tests of the RYD-Box \cite{ryd_box} and the Volkswagen (VW) Data Plug \cite{data_plug} in this paper. We ran two different test cases both were carried out with a Volkswagen Golf model 7: In case $T1$ messages are blocked (Table \ref{tab:blocked_can_messages} contains detailed information about the blocked messages). In case $T2$, we manipulate messages' contents (for details see Table \ref{tab:manipulated_can_messages}). In addition to the respective PID (process identifier) of the message, we have also included the minimum and maximum values defined in the standard and the firewall response defined by our rule in the tables, e.g. \texttt{blocked} in case $T1$ and limiting the speed value to a maximum of \texttt{100} in $T2$. Finally, we also theoretically discuss which threats  identified using STRIDE and sorted using DREAD can be mitigated by our approach.

\subsection{Testing the RYD-Box}%
\label{testing_ryd_box}
	First we ran $T1$ and blocked CAN requests for both speed and current mileage using our firewall. Then we did a test drive and checked the data displayed to us in the app. Here we were still shown the last known mileage, which no longer matched after the drive with the firewall activated. We had tested the dongle beforehand without the firewall, so the service already had access to our mileage at that time. This shows that the mileage is stored in the cloud at certain times of the respective car, but also that it is possible to successfully block the transmission of the mileage with our approach and still use the remaining functions of the dongle. We then displayed the speed history of our recorded journey. There we noticed that despite blocking the messages for the current speed on the OBD-II interface, the approximate driving speed was also tracked in the app. Our first assumption was that the app calculated the speed using the GPS data of the smartphone. However, after another drive without a smartphone in the car, we found out that the RYD-Box has both a GSM module with a separate SIM card and a GPS sensor. This enables the service to approximate the speed using GPS. Alternatively, additional sensors such as an acceleration sensor or a gyroscope could be installed to collect driving statistics. However, we have not yet disassembled the dongle and can therefore only list additional possibilities to the GSM and GPS modules.
	
	Afterwards, we reconfigured our firewall for $T2$ so that it always returns 200000 km when the current mileage is requested. In addition, we have limited the current speed to 50 km/h in a rule using behaviour. Then we did the same test drive as for $T1$ and compared the results in the app again. As we expected, the speed limit had no effect on our tracked distance. We simply assume that the application does not use the current speed transmitted via CAN for the calculations or displays within the app for a journey. However, we could see modified mileage, namely the value of 200000 km as defined in the behaviour of our rule, becoming displayed in RYD's smartphone app, without any hint on it being modified. 

\subsection{Testing the VW Data Plug}
\label{testing_vw_data_plug}
	We blocked the CAN requests to the OBD-II gateway for $T1$ using our firewall for the current mileage and the current speed. After connecting our mobile phone via bluetooth to the dongle and the app, we were able to display data from our vehicle. However, the mileage could not be displayed because it was successfully blocked by our firewall. Once again, we drove a test lap for the recording of a journey. Again, just like with the RYD-Box, a speed could be read out in the app afterwards with the Volkswagen Data Plug, despite the blocking by our firewall. However, here again we have the same case as in the test with the RYD-Box. The speed for the saved journeys is simply not displayed over the speed that can be requested via OBD-II. In contrast to the RYD-Box, the VW Data Plug from Volkswagen does not have a GSM or GPS module, but the application must have access to the current location and the GPS services of the smartphone in order to function at all. For this reason, no test was possible without a smartphone in the vehicle. Due to the compact and smaller design of the VW Data Plug, we can assume that no additional sensors are installed there, and only the sensors installed in the smartphone (such as the acceleration sensor or the gyroscope) are used to evaluate the driving behaviour.
	
	For the $T2$, our firewall was again configured in the same way as it was above during the test for the RYD-Box. After our new round of tests, we were able to prove exactly the same behaviour as the RYD-Box showed. We were able to see our manipulated vehicle mileage within the We Connect application, but the manipulation of the speed had no effect, as the speed is approximated by GPS as explained before.
\vspace{-2em}
\begin{table}[H]
	\centering
	\resizebox{0.8\textwidth}{!}{%
		\begin{tabular}{|l|c|c|c|c|}
			\hline
			\textbf{PID (hex)}    & \textbf{min\_value} & \textbf{max\_value} & \textbf{Firewall response} & \textbf{Description}                     \\ \hline
			0C & 0 & 255 & $<blocked>$ & Vehicle speed \\ \hline
			A6 & 0 & 429,496,729.5 & $<blocked>$ & Odometer \\ \hline
		\end{tabular}%
	}
	\caption{Overview of CAN-Messages to be blocked in $T1$}
	\label{tab:blocked_can_messages}
\end{table}
\vspace{-3em}
\begin{table}[H]
	\centering
	\resizebox{0.8\textwidth}{!}{%
		\begin{tabular}{|l|c|c|c|c|}
			\hline
			\textbf{PID (hex)}    & \textbf{min\_value} & \textbf{max\_value} & \textbf{Firewall response} & \textbf{Description}                     \\ \hline
			0C & 0 & 255 & $<min\_value>$ to 100  & Vehicle speed \\ \hline
			A6 & 0 & 429,496,729.5 & 200,000.0 & Odometer \\ \hline
		\end{tabular}%
	}
	\caption{Overview of CAN-Messages to be manipulated in $T2$}
	\label{tab:manipulated_can_messages}
\end{table}
 
\subsection{Evaluation of Existing Dongles }
\label{conclusion_of_testing}
\vspace{-0.2em}
	These two examples in Section \ref{testing_ryd_box} and \ref{testing_vw_data_plug} are only intended to show the feasibility and do by no means represent a full analysis of all car manufacturers in combination with different dongles. Because our approach is protocol agnostic and also not dependent on the underlying message format, it can be assumed that our OBD-II firewall will work with all possible combinations. Our main takeaway from testing against real existing dongles is that our approach is
	surprisingly easy to apply in practice.
	This means that no message origin authentication features (like digital signatures on message contents \cite{DBLP:conf/camad/PohlsP17,DBLP:conf/icics/0003SPF16}) on the CAN bus are used, as the dongle does not detect the manipulation of message's content in test case $T2$.
	
	Additionally we gained interesting deeper insights into the functionality of the two dongles. 
	We contemplate a closer look at the dongles in combination with the related smartphone application, such as in the case of the Volkswagen Data Plug, because we found that this combined setup additionally used information collected from the smartphone's sensors; these might also be tampered with - albeit in different ways:
	We consider it quite conceivable to trick the application corresponding to the dongle to the extent of feeding it false journeys with faked speeds. To do this, you would have to isolate the application to run it within an emulator where you can  influence the position information and the gyroscope's readings.

\subsection{Evaluation of Threat Mitigation}
\label{threat_impacts_with_firewall}
Finally we check whether our approach prevents or at least mitigates 
the threats $T_\alpha$ to $T_\eta$ as identified in Section  \ref{sec:threat_identification}.  
With regard to the threats $T_\alpha$ and $T_{\beta_I}$, our approach unfortunately cannot do anything. In general, it is almost impossible to achieve hundred percent security against physical attacks or manipulations. With regard to the $\beta$ threat class, however, you can try to make your services within the vehicle as secure as possible and close any security gaps discovered with regular updates. However, if the attacker has physical access, often even the best systems cannot be protected. Direct and unlimited access to low-level debug interfaces such as those provided by JTAG (Joint Test Action Group) and SWD (Serial Wire Debug) would make it possible to take complete control of the device, e.g. stop and change code execution, access memory and registers, or even dump the firmware.

However, $T_\gamma$ is preventable by means of our approach. With the help of the rules and the behaviours, it can be defined within the firewall which messages are allowed through and which are not. This means that it is no longer possible for an attacker to simply send arbitrary messages.

Likewise, the threat $T_\delta$  can be prevented by configuring the firewall for the desired permitted communication using the method described above. 
Our approach not only allows or blocks certain messages, but also enables to modify the data payload for defined messages before publishing them to the OBD-II device. 

Since the threats $T_\gamma$ and $T_\delta$  can be prevented by our approach, the threats based on $T_\gamma$ can also be avoided automatically. Thus, threats $T_\epsilon$, $T_\zeta$ and $T_\eta$ are also eliminated. This means that our firewall approach successfully prevents all threats on the hierarchies $L_{II}$ and $L_{III}$ (see Figure \ref{fig:threats_attack_tree}).

\section{Conclusion}
\label{conclusion}
While it is yet not really widespread to install filtering approaches such as firewalls in vehicle systems at the OBD-II outgoing interface, this can become a crucial interconnection point between the car and third party dongles in the future.
The need for such solutions will grow in the coming years, as more and more different approaches and products will come onto the market to make existing cars even smarter. And the OBD-II interface is clearly the standard entry point for realising such smart car approaches.
Manufacturers agree that security must be added to CAN or OBD-II, but dongles available on the market today can be easily fooled by manipulated messages. 
%
%
Following the threats we have systematically defined in Section ~\ref{sec:threat_identification}, the proposed Man-in-the-OBD shows that without any means to authenticate messages from the car's electronic control units, e.g. the car's current speed, dongles connected to the OBD-II interface can be fooled.
We are the first to show that a Man-in-the-OBD-II interface is able to serve the dongle manipulated values without it noticing the falsification.
We exploit the missing authentication to protect the car's drivers privacy.
This is of course a benign application, 
 used maliciously the falsified information provided to the environment of the car could have negative consequences. During the test cases, we were also able to test whether our firewall without rules has an effect on the functionality of the dongles used. This was not the case which can be easily explained by the fact that our approach in normal cases without active roles is simply two physically separated CAN-BUS systems forwarding all messages between the buses. 

$L_{I}$-threats are directly attacking the dongle, like $T_\alpha$ and $T_{\beta_I}$.
Those can of course not be mitigated by our firewall,
because our firewall simply is a dongle.
However, we show that threats of hierarchies $L_{II}$ and $L_{III}$ can be effectively prevented by policing the OBD-II interface. 
This is already an important step to protect against external malicious devices. 
We have implemented it and prototypically showed that it works for the CAN protocol. 
However, our architecture is general, and our approach could be used even for other interfaces; our modular approach allows to extend it to other protocols. 

For an increased privacy we showed that the connected dongles providing data for cloud services to monitor the car via a mobile phone application are not affected by the small delay caused by our  filtering or manipulation of the data traffic.
%

\subsection{Future Work}
\label{future_work}
%
Due to the diversity of protocols available via OBD-II, we limited ourselves to the CAN protocol. It would definitely be an enrichment for our firewall if the remaining protocols accessible via OBD-II were also covered by our firewall. 
%

Furthermore, the OBD-II interface is the most exposed interface, but it would be valuable to research 
a firewall or man-in-the-middle attack at 
other interfaces; for example the infotainment system. 
Technologies such as AndroidAuto \cite{android_auto} and Apple CarPlay \cite{apple_carplay} are already finding their way into vehicles. But also completely independent Android systems are delivered with the so-called Android Automotive OS \cite{android_os}. 
If apps would have vulnerabilities or contain malicious code 
it would be fatal if such apps could suddenly gain control over the vehicle. Therefore, an adapted firewall approach similar to our current one could also be very useful to isolate the car from the infotainment system. Another point that should be examined in future work is how to implement our developed approach on less expensive devices than the Raspberry Pi. If applicable, approaches should also be considered here that shift the actual computations to a central processing unit provided by the car, as an example. 



  

%
%
%
 \bibliographystyle{splncs04}
 \bibliography{fk}

\end{document}